\documentclass[aps,preprint,groupedaddress,showpacs,longbibliography,nofootinbib]{revtex4-1}
\usepackage{amssymb}
\usepackage[breaklinks,colorlinks,citecolor=green,urlcolor=blue,linkcolor=red]{hyperref}
\usepackage{graphicx}
\usepackage{epstopdf}
\usepackage{diagbox}
\usepackage{float}
\usepackage{subfigure}
\usepackage{multirow}
\usepackage{CJK}

\begin{document}

%\preprint{}

\title{\boldmath Analysis of angular distribution asymmetries and the associated $C\!P$ asymmetries in three-body decays of bottom baryons}


\author{Zhen-Hua Zhang}
\email{zhangzh@usc.edu.cn}
%\thanks{}
\affiliation{School of Nuclear Science and Technology, University of South China, Hengyang, 421001, Hunan, China}
\author{Jing-Juan Qi}
\email{jjqi@mail.bnu.edu.cn}
\affiliation{College of Information and Intelligence Engineering, Zhejiang Wanli University, Zhejiang 315101, China}


\date{\today}

\begin{abstract}
We introduce a set of observables representing angular distribution asymmetries, which can be viewed as a generalization of the forward-backward asymmetry of angular distributions, and can be used as an effective tool to search for $C\!P$ violation in three-body decays of bottom and charmed baryons.
We propose to search for such $C\!P$ asymmetries (1) in decays with $\Lambda^0$, $\Sigma^\pm$, or $\Lambda_c^+$ involved, such as $\Lambda_b\to \Lambda^0 D$ and $\Lambda_b\to\Lambda^0\rho(770)^0$; and (2) in three-body decays of bottom baryons with opposite parity intermediate resonances involved.
Typical examples include $\Xi_b^-\to p K^- K^-$, $\Lambda_b^0\to p K_S\pi^-$ and $\Lambda_b^0\to p \pi^0\pi^-$, in which the last decay channel is used as a toy model to illustrate of the basic idea.
\end{abstract}

%\pacs{}
%\keywords{CP violation, heavy baryon,}

\maketitle

%\section{}
%\subsection{}
%\subsubsection{}

\section{Introduction}
$C\!P$ violation ($C\!P$V), which is an important component of the Standard Model (SM) of particle physics \cite{Kobayashi:1973fv}, has been observed in $K$, $B$, $B_s$ and $D$ meson decay processes.
Theoretical investigations of $C\!P$V have been performed in baryon decay processes\cite{Giri:2001ju,Leitner:2006sc,Lu:2009cm,Kang:2010td,Hsiao:2014mua,Gronau:2015gha,Durieux:2016nqr,Bigi:2017eni,Shi:2019vus,Sinha:2021mmx,Zhang:2021sit,Geng:2022osc,Wang:2022tcm}. It has been shown that there can be relatively large $C\!P$ asymmetries ($C\!P$As) in some decays of bottom baryons \cite{Hsiao:2014mua}.
On the experimental side, the $C\!P$As have been investigated in some typical two-body decay channels of $\Lambda_b^0$ by the Collider Detector at Fermilab (CDF) \cite{CDF:2011ubb,CDF:2014pzb} and  Large Hadron Collider beauty (LHCb) \cite{LHCb:2018fly}, in three- or four-body decays of $\Lambda_b^0$, $\Xi_b^0$, $\Xi_b^-$, $\Lambda_c^+$ and $\Xi_c^+$ by LHCb \cite{LHCb:2016yco, LHCb:2018fpt,LHCb:2019jyj,LHCb:2019oke,LHCb:2021enr,LHCb:2017hwf,LHCb:2020zkk}, and very recently in hyperon decays by the Beijing Spectrometer III (BESIII) \cite{BESIII:2022qax} and Belle \cite{Belle:2022uod} experiments.
However, the pursuit of $C\!P$V in the baryonic sectors, which is of great importance for testing the SM and for the indirect search for new physics beyond SM, has not had a positive result after years of efforts.

Since baryons are particles with spins, other than the partial decay width, $C\!P$V in the baryon decay processes can be present in observables associated with the angular distributions of the final particles.
One such observable is the decay parameter in two-body weak decay processes of baryons.
Typical examples include $\Lambda^0\to p\pi^-$ and $\Lambda_b^0\to D\Lambda^0$, where the decay parameter and the associated $C\!P$As of the former channel have been measured by BESIII through baryon-anti-baryon pair production $J/\psi\to\Lambda^0\overline{\Lambda^0}$ \cite{BESIII:2022qax}, and those of the latter were proposed for extracting the Cabibbo-Kobayashi-Maskawa (CKM) phases $\gamma$ through the dual weak cascade decays $\Lambda_b^0\to D\Lambda^0$ with $\Lambda^0\to p\pi^-$ \cite{Giri:2001ju,Zhang:2021sit,Geng:2022osc}.
In both cases, the decay parameters can be related to forward-backward asymmetries (FBA) of the final particle distributions in certain reference frames.

$C\!P$V can also leave tracks in the angular distributions of final particles in multi-body decay processes of hadrons.
Examples include $C\!P$As associated with the triple product asymmetries (TPAs) in baryon decay processes \cite{Bensalem:2002ys,Gronau:2015gha,LHCb:2018fpt,LHCb:2019oke} and partial-wave $C\!P$As (PW$C\!P$As) \cite{Zhang:2021fdd}.
In fact, the largest $C\!P$As ever observed are those localized in certain regions of the phase space in three-body decay channels of bottom mesons, such as $B^\pm\to\pi^\pm\pi^+\pi^-$, $B^\pm\to K^\pm K^+ K^-$, $B^\pm\to K^+\pm \pi^+\pi^-$ and $B^\pm\to \pi^\pm K^+ K^-$ \cite{LHCb:2013ptu,LHCb:2013lcl,LHCb:2014mir,LHCb:2019xmb,LHCb:2019jta,LHCb:2019sus}. %, though the integrated $C\!P$As are all relatively quite small \cite{LHCb:2013ptu,LHCb:2013lcl,LHCb:2014mir,LHCb:2019xmb,LHCb:2019jta,LHCb:2019sus}.
Take $B^\pm\to\pi^\pm\pi^+\pi^-$ as an example. %, in which
Very large regional $C\!P$As were observed in part of the $f_0(500)-\rho(770)^0$ interference region corresponding to the angle between the two same-sign pions smaller than $90^\circ$.
This large regional $C\!P$A can be explained by the interference of the $s$- and $p$-wave amplitudes (corresponding to $f_0(500)$ and $\rho(770)^0$, respectively) with a natural inclusion of a non-perturbative strong phase difference between the two waves \cite{Zhang:2013oqa}, which can be ideally studied through the angular distribution asymmetry observables FBA \cite{Zhang:2021zhr,Wei:2022zuf}. %induced $C\!P$As (FB-$C\!P$As)
It is naturally expected that there are similar $C\!P$As associated with the anisotropy of angular distributions of the final particle in multi-body decays of baryons.

The aim of this paper is to introduce a set of angular distribution asymmetry observables which can be viewed as a generalization of the aforementioned FBAs in the multi-body decay processes of baryons and mesons.
The newly introduced observables can be used in searching for $C\!P$ violations in the baryon decay processes, especially in the bottom baryon decays.

The remainder of this paper is organized as follows.
In Sect. \ref{sec:ADA}, we present the definition of the angular distribution asymmetry observables and their corresponding $C\!P$V observables.
In Sect. \ref{sec:app}, we discuss the potential applications of the newly introduced $C\!P$V observables.
In Sect. \ref{sec:toymode}, we use a toy model, $\Lambda_b^0\to p\pi^0\pi^-$, to illustrate the basic idea.
In the last section, we give a brief summary of this paper.

\section{\label{sec:ADA} \boldmath angular distribution asymmetry and the corresponding $C\!P$V observables}

\begin{figure}
  \centering
  \includegraphics[width=1.0\textwidth]{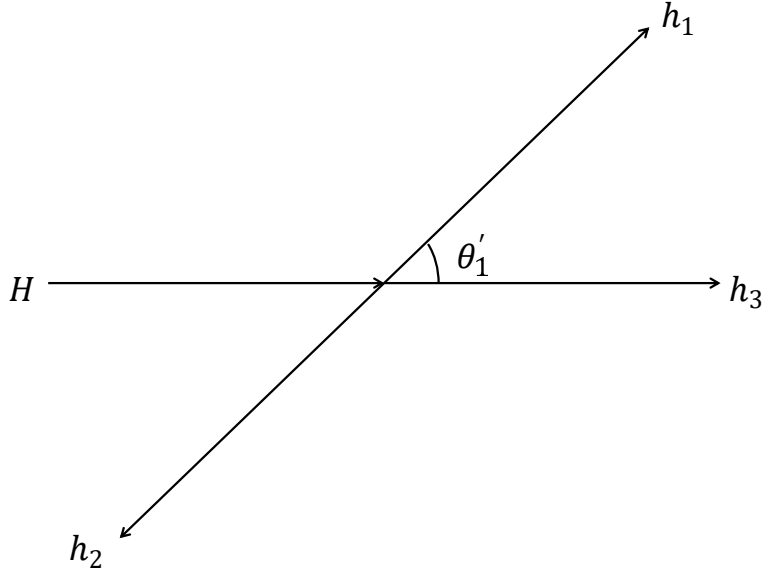}\\
  \caption{Illustration of the definition of $\theta_1'$.}\label{fig:theta}
\end{figure}

To put the discussion on more general grounds, we consider a three-body weak-decay process of a hadron $H$, $H\to  h_1h_2h_3$.
It can be proven that the square of the spin-averaged decay amplitude, which is defined as $\overline{\left|\mathcal{M}^{J}\right|^2}\equiv\frac{1}{2J+1}\sum_{m_z,\lambda_k}\left|\mathcal{M}_{\lambda_1\lambda_2\lambda_3}^{Jm_z}\right|^2
$ for unpolarized $H$, where $\mathcal{M}_{\lambda_1\lambda_2\lambda_3}^{Jm_z}$ is the corresponding decay amplitude in the helicity form,
can be expressed as
\begin{equation}\label{MJ}
\overline{\left|\mathcal{M}^{J}\right|^2}=\sum_j w^{(j)} P_j\left(c_{\theta_1'}\right) ,
\end{equation}
where $J$ and $m_z$ are the spin and its $z$-axis component of $H$, $\lambda_i$ ($i=1,2,3$) is the helicity of $h_i$, $P_j$ is the $j$-th Legendre polynomial, $w^{(j)}$ represents the weight of the $j$-th wave, $c_{\theta_1'}\equiv\cos(\theta_1')=\frac{s_{13}-(s_{13,\text{max}}+s_{13,\text{min}})/2}{(s_{13,\text{max}}-s_{13,\text{min}})/2}$, where $\theta_1'$ is the angle between the momenta of $h_1$ and $H$ in the center-of-mass (c. m.) frame of the $h_1h_2$ system (see Fig. \ref{fig:theta} for illustration; note that $\theta_1'$ is also the angle between the momenta of $h_1$ and $h_3$ in the same frame), and $s_{13,\text{min}}$ and $s_{13,\text{max}}$ are the minimum and the maximum of the $s_{13}$ constraint by the phase space.

The presence of odd-$j$ terms $w^{(j)}$ results in angular distributions asymmetries, i.e., asymmetries between $\theta_1'\leftrightarrow \pi-\theta_1'$.
To account for this kind of asymmetry, we introduce a set of observables, which is defined as
\begin{equation}\label{APL}
  A^{FB}_{j}= \frac{\left(-\int_{-1}^{x_{1}^{(j)}}+\int_{x_{1}^{(j)}}^{x_{2}^{(j)}}-\int_{x_{2}^{(j)}}^{x_{3}^{(j)}}\cdots +\int_{x_j^{(j)}}^{+1}\right)\overline{\left|\mathcal{M}^{J}\right|^2} dc_{\theta_1'}}{\int_{-1}^{+1}\overline{\left|\mathcal{M}^{J}\right|^2} dc_{\theta_1'}}
\end{equation}
for odd $j$, where $x_k^{(j)}$ ($k=1,2,\ldots,j$) is the $k$-th zero point of the Legendre polynomial $P_j(x)$.
Note that this can be viewed as a generalization of the FBA for meson decays such as $B^\pm\to\pi^\pm\pi^+\pi^-$, or for baryon decays such as $\Lambda_b^0\to \Lambda(\to p \pi^-)D$.\footnote{Note, however, that the newly introduced observables cannot be viewed as a generalization of the decay parameter of the hyperon two-body decay $\Lambda^0\to p \pi^-$ \cite{Lee:1957he,Lee:1957qs}.
As the decay parameter of the hyperon decay $\alpha^{\Lambda\to p\pi^-}$ is defined according to the angle between the spin of the hyperon and the momentum of the proton, which is a parity-odd quantity, the corresponding $C\!P$A can be defined as $A_{C\!P}^{\alpha^\Lambda}\equiv \frac{1}{2}\left(\alpha^{\Lambda\to p\pi^-}+\alpha^{\overline{\Lambda}\to \overline{p}\pi^+}\right)$.}
As one can see for the case $k=1$, $A^{FB}_{1}$ reduces to the FBA.
Hence, we will call $ A^{FB}_{j}$ the $j$-th FBA.
The corresponding $C\!P$-violating observables, which will be named the $j$-th FBA-induced $C\!P$ asymmetry ($j$-th FB-$C\!P$A), can then be defined as
\begin{equation}\label{AFBCPj}
   A^{FB}_{j,C\!P}=\frac{1}{2}\left(A^{FB}_{j}-\overline{A^{FB}_{j}}\right),
\end{equation}
where $\overline{A^{FB}_{j}}$ is the $j$-th FBA of the charge conjugation process, and the presence of
the minus sign is because $A^{FB}_{j,C\!P}$ and $\overline{A^{FB}_{j}}$ are parity-even observables.\footnote{
$C\!P$V can be investigated through PW$C\!P$A, which are defined as $A_{CP}^{(j)}\equiv \frac{w^{j}-\overline{w^{j}}}{w^{j}+\overline{w^{j}}}$.
However, in order to get $w^{j}$ and $\overline{w^{j}}$, partial-wave analysis (PWA) is inevitable.
The events distributed in different locations of the phase space are not treated equally in PWA,
which complicates the experimental analysis.
As an example of PWA, see Ref. \cite{BaBar:2012iuj}.}

One can easily see from the definition that $A^{FB}_{j}$ (and hence $A^{FB}_{j,C\!P}$) can only get contributions from $w^{(j')}$ with odd $j'$.
None of the $w^{(j')}$s with even $j'$ contribute to any of the $A^{FB}_{j}$s.
On the other hand, there is no one-to-one correspondence between $A^{FB}_{j}$ and $w^{(j)}$, meaning that each $A^{FB}_{j}$ obtains contributions from all the $w^{(j')}$ for odd $j'$.
Despite of this, for a fixed $j=j_0$, the most important contributions to $A^{FB}_{j_0}$ and $A^{FB}_{j_0,C\!P}$ come from $w^{(j_0)}$.

In what follows, we consider cascade decay $H\to  R_i (\to h_1 h_2) h_3$, where the subscript $i$ in $R_i$ indicates that there may be more than one intermediate particle with similar masses.
We first leave it open to whether $R_i\to h_1 h_2$ is weak or strong.
$w^{(j)}$ can be expressed in terms of the corresponding decay amplitudes.
After some algebra, one has
\begin{equation}\label{wj}
  w^{(j)}=\sum_{ii'}\left\langle\frac{\mathcal{S}_{ii'}^{(j)}\mathcal{W}_{ii'}^{(j)}}{\mathcal{I}_{R_i}\mathcal{I}_{R_{i'}}}\right\rangle,
\end{equation}
where the notation ``$\langle\cdots\rangle$'' indicates the integral with respect to $s_{12}$ over a small interval which covers all the resonances of interest $R_i$, and $\mathcal{W}_{ii'}^{(j)}$ and $\mathcal{S}_{ii'}^{(j)}$ contain the decay amplitudes of $H\to R_i h_3$ and $R_i\to h_1h_2$, respectively, and take the form
\footnote{One can also replace the helicity decay amplitudes by the spin-angular ones. For example,
$\mathcal{F}^{J}_{R_i,\sigma\lambda_3}=\sum_{ls} \left(\frac{2\,l+1}{2J+1}\right)^{\frac{1}{2}}\langle l0s\sigma-\lambda_3|lsJ\sigma-\lambda_3\rangle \langle s_{R_i}\sigma s_3\lambda_3|s_{R_i}s_3s \sigma-\lambda_3\rangle a^J_{R_{i},ls}$.
}
\begin{equation}\label{Wj}
  \mathcal{W}_{ii'}^{(j)}=\sum_{\sigma\lambda_3}(-)^{\sigma-s}\langle s_{R_i}-\sigma s_{R_{i'}}\sigma|s_{R_i}s_{R_{i'}}j0\rangle\mathcal{F}^{J}_{R_i,\sigma\lambda_3}\mathcal{F}^{J\ast}_{R_{i'},\sigma\lambda_3},
\end{equation}
and
\begin{equation}\label{Sj}
  \mathcal{S}_{ii'}^{(j)}=\left.\sum_{\lambda_1'\lambda_2'}(-)^{s-\lambda'}\langle s_{R_i}-\lambda's_{R_{i'}}\lambda'|s_{R_i}s_{R_{i'}}j0\rangle \mathcal{F}^{R_i,s_{R_i}}_{\lambda_1'\lambda_2'}\mathcal{F}^{R_{i'},s_{R_{i'}}\ast}_{\lambda_1'\lambda_2'}\right|_{\lambda'=\lambda_1'-\lambda_2'},
\end{equation}
where the notation $\langle \cdots|\cdots\rangle$s are the Clebsch-Gordan coefficients.
Note that the introduction of $s$ in the above two equations is to make $\sigma-s$ and $s-\lambda'$ integers and can take the form of either $s_{R_i}$ or $s_{R_{i'}}$.
The Clebsch-Gordan coefficients in Eqs. (\ref{Wj}) and (\ref{Sj}) restrict $j$ in Eq. (\ref{MJ}) so that it can
only take integer values from 0 to $\max_i (2s_{R_i})$.

\section{\label{sec:app}Applications}
%{\it Application.---}
For the applications of the newly introduced observables $A^{FB}_{j}$ and $A^{FB}_{j,C\!P}$, we want to consider two situations, according to whether the decay $R_i\to h_1h_2$ is weak or strong.
In the first one, the decay $R_{i^{(\prime)}}\to h_1h_2$ is a weak process.
In this situation, the intermediate state $R_{i^{(\prime)}}$ has negligibly narrow decay width.
Hence, there is no need to consider the interference of nearby resonances, which means that there is only one intermediate state, $R_i$.
Typical processes include (1) $\Lambda_b^0\to \Lambda^0 M$, with $M$ being mesons such as $\pi$, $\rho$, $D$, or $J/\psi$, and $\Lambda^0$ as $R_i$ and decaying through $N\pi$; (2) $\Lambda_b^0\to \Sigma^\pm M$, with $\Sigma^\pm$ as $R_i$ and decaying through $N\pi$; (3) $\Lambda_b^0\to \Lambda_c^+ M$, with $\Lambda_c^+$ as $R_i$ and decaying through $p K_s$ or $\Lambda^0\pi^+$.
Since the sub-process $R_{i^{(\prime)}}\to h_1h_2$ is a weak process, there is no extra constraint from the parity conservation.
The only constraint worth mentioning comes from the Clebsch-Gordan coefficients in Eqs. (\ref{Wj}) and (\ref{Sj}), which tells us that $j$ can only take integer values  0 and 1 for all these aforementioned examples with spin-parity $\left(\frac{1}{2}\right)^+$ baryons as $R_i$.
Hence, the square of the spin-averaged amplitude always takes the form
$\overline{\left|\mathcal{M}^{J}\right|^2}\propto 1+A^{FB}_1 c_{\theta_1'}$.
The only practically relevant observable is the first FBA $A^{FB}_1$, which is in fact the well known decay asymmetry parameter.

Another situation is when the decays $R_{i^{(\prime)}}\to h_1h_2$ are strong processes.
These strong decay processes respect parity symmetry, which implies from Eq. (\ref{Sj}) that
\begin{equation}\label{SR}
  \mathcal{S}_{ii'}^{(j)}=\Pi_{R_i}\Pi_{R_{i'}}(-)^j\mathcal{S}_{ii'}^{(j)},
\end{equation}
where $\Pi_{R_{i^{(\prime)}}}$ is the parity of $R_{i^{(\prime)}}$.
If there is only one resonance $R_i$ entering into the decay process, Eq. (\ref{SR}) will reduce to $\mathcal{S}_{ii}^{(j)}=(-)^j\mathcal{S}_{ii}^{(j)}$.
Hence $\mathcal{S}_{ii}^{(j)}=0$ for odd $j$.
All the $w^{(j)}$ will simply be zero for odd j!
This means that there is no need for the consideration of $A^{FB}_j$ and $A^{FB}_{j,C\!P}$ at all.
The only way out of this dilemma is when there are (at least) two resonances, say $R_{i_1}$ and $R_{i_2}$, with opposite parities and similar masses.
It is this situation that has the most similarities with the aforementioned three-body decays of $B^\pm$, in which the largest regional $C\!P$As are observed.
Now $\mathcal{S}_{i_1i_2}^{(j)}$ can be nonzero for odd $j$ according to Eq. (\ref{SR}); hence, $A^{FB}_j$ and $A^{FB}_{j,C\!P}$ can be nonzero.

The interference of nearby intermediate resonances are fairly common phenomena in multi-body decay of bottom or charmed hadrons.
We list two typical examples of three-body decay of bottom baryons for the second situation.

The first one is $\Xi_b^-\to p K^- K^-$.
It has already been observed by LHCb that there are some resonant structures in the low invariant mass region of the $p K^-$ system, such as $\Sigma(1775)$ and $\Sigma(1915)$, whose spin-parities are $(\frac{5}{2})^+$ and $(\frac{5}{2})^-$, respectively \cite{LHCb:2021enr}.
Consequently, $j$ can take integer values from 0 to 5, according to Eqs. (\ref{Wj}) and (\ref{Sj}).
Hence, $w^{(1)}$, $w^{(3)}$ and $w^{(5)}$ will be nonzero.
Correspondingly, there can be angular distribution asymmetries, which are suitably studied through the measurements of $A^{FB}_{1}$, $A^{FB}_3$ and $A^{FB}_5$.
Moreover, the associated $C\!P$ asymmetries, $A^{FB}_{1,C\!P}$, $A^{FB}_{3,C\!P}$, and $A^{FB}_{5,C\!P}$, can also be measured.
It should be pointed out that $A^{FB}_3$ and $A^{FB}_5$  (hence $A^{FB}_{3,C\!P}$ and $A^{FB}_{5,C\!P}$) have never been studied in any decay channels before.

Anther example is $\Lambda_b^0\to p K^*(892)^0\pi^-$.
Although the nature of the resonant structures observed by LHCb in the low invariant mass region of the $p \pi^-$ system remains unclear \cite{LHCb:2014yin}, there is still a good chance for the presence of $N^+(1440)$ and $N^+(1520)$, whose spin-parities are $(\frac{1}{2})^+$ and $(\frac{3}{2})^-$, respectively, and $j$ can take integer values 0, 1, and 2, according to Eqs. (\ref{Wj}) and (\ref{Sj}).
Consequently, $w^{(1)}$ will be nonzero.
The corresponding angular distribution asymmetries, $A^{FB}_{1}$, will be nonzero.
There can also be nonzero associated $C\!P$ asymmetries, $A^{FB}_{1,C\!P}$.

\section{\label{sec:toymode}\boldmath $C\!P$V analysis for $\Lambda_b^0\to p \pi^0 \pi^-$}
To illustrate, we present in this section a brief $C\!P$V analysis of the decay channel $\Lambda_b^0\to p \pi^0\pi^-$.
We consider the interference effects of the intermediate resonances $N(1440)^+$ and $N(1530)^+$, whose spin-parities are ${\frac{1}{2}}^+$ and ${\frac{3}{2}}^-$, respectively.
When the square of the invariant mass of the $p \pi^0$ system, $s$, is around the masses of $N(1440)^+$ and $N(1530)^+$,
the decay is dominated by the cascade decays $\Lambda_b^0\to N_j(\to p \pi^0) \pi^-$ (with $N_1$ and $N_2$ representing $N(1440)^+$ and $N(1530)^+$, respectively).
Hence, the decay amplitudes can be expressed as a summation of those corresponding to the two aforementioned baryonic resonances, which reads
\begin{equation}\label{eq:amp}
  \mathcal{M}=\mathcal{M}_{N_1}+\mathcal{M}_{N_2}e^{i\delta}, %e^{i\delta},
\end{equation}
where $\mathcal{M}_{N_j}$ ($j=1,2$) represents the amplitudes for the cascade decay $\Lambda_b^0\to N_j(\to p \pi^0) \pi^-$,
and $N_{1}$ and $N_{2}$ represent $N(1440)^+$ and $N(1530)^+$, respectively, which take the form
\begin{equation}%\label{}
  \mathcal{M}_{N_j}=\frac{1}{s_{N_j}}\sum_{\text{pol.~} N_j}\mathcal{M}_{\Lambda_b\to N_j\pi^-}\mathcal{M}_{N_j\to p\pi^0},
\end{equation}
where $\mathcal{M}_{\Lambda_b\to N_j\pi^-}$ and $\mathcal{M}_{N_j\to p\pi^0}$ are the decay amplitudes for the two-body processes $\Lambda_b\to N_j\pi^-$ and $N_j\to p\pi^0$, respectively, and $s_{N_j}=s-m_{N_j}^2+im_{N_j}\Gamma_{N_j}$.
Note that a strong phase $\delta$ is introduced in Eq. (\ref{eq:amp}).
This strong phase is in fact the phase difference between the effective strong couplings of $N_1\to p\pi^0$ and $N_2 \to p \pi^0$.

After some algebra, the square of the spin-averaged decay amplitude of the decay $\Lambda_b^0\to p \pi^0\pi^-$ can be expressed as
\begin{equation}\label{eq:Msqr}
  \overline{|\mathcal{M}|^2}=\frac{\lambda(m_{\Lambda_b}^2,s,0)\lambda(m_p^2,s,0)}{s}
  \left[\frac{|\alpha_{N_1}|^2}{|s_{N_1}|^2}+\frac{|\alpha_{N_2}|^2}{|s_{N_2}|^2}\left(1+3\cos^2\theta\right)
  +12\Re\left(\frac{\alpha_{N_1}\alpha_{N_2}^\ast e^{i\delta}}{s_{N_1}s_{N_2}^\ast}\right)\cos\theta\right],
\end{equation}
where the kinematic function $\lambda$ is defined as $\lambda(a,b,c)\equiv a^2+b^2+c^2-2ab-2bc-2ca$, %$s$ is the invariant mass squared of the $p \pi^0$ system, $s_{R}=s-m_{R}^2+im_{R}\Gamma_{R}$,
and
$\alpha_{N_j}\propto\lambda_{u}a_{N_j}^{\text{tree}}-\lambda_{t}a_{N_j}^{\text{penguin}}$, with the CKM factors taking the form $\lambda_q=V_{qb}V_{qd}^\ast$, and $a_{N_j}^{\text{tree}}$ and $a_{N_j}^{\text{penguin}}$ representing the remaining factors of the tree and penguin parts for the decay processes $\Lambda_b^0\to N_j(\to p \pi^0) \pi^-$.
The angle $\theta$ is now the relative angle between the momenta of the two pions in the c.m. frame of the $p \pi^0$ system.
In deriving the above expression, we have taken the limit $m_\pi\to 0$.
In addition, we have replaced $m_{N_1}^2$ and $m_{N_2}^2$ by $s$ except in the Breit-Wigner factors $1/s_{N_j}$.
This is reasonable because we are working in a small region of the phase space around $N(1440)^+$ and $N(1530)^+$; hence, $s$ is close to $m_{N_j}$.

The decay amplitudes for the $C\!P$-conjugate process $\overline{\Lambda_b^0}\to \overline{p} \pi^+ \pi^0$ can be obtained by replacing the CKM matrix elements in Eq. (\ref{eq:Msqr}) by their complex conjugates.
The $C\!P$V behavior is now clear to us.
The first two terms in Eq. (\ref{eq:Msqr}), which are even under the exchange $\theta\leftrightarrow \pi-\theta$, contain the $C\!P$V of the two-body decays $\Lambda_b^0\to N_1\pi^-$ and $\Lambda_b^0\to N_2\pi^-$, respectively.
This $C\!P$V is caused by the interference of $a_{N_j}^{\text{tree}}$ and $a_{N_j}^{\text{tree}}$, while the last term represents the interference between $N_1$ and $N_2$.
One can easily see from the definition of the FBA $A_{1}^{FB}$ in Eq. (\ref{APL}) that only this term is present in the numerator of $A_{1}^{FB}$, since this term is proportional to $\cos\theta$ and hence is odd under the exchange $\theta\leftrightarrow \pi-\theta$.
The presence of the strong phase $\delta$ in this term is crucial to the corresponding FB-$C\!P$A $A_{1,CP}^{FB}$.
It is possible that for proper values of $\delta$ realized in nature,
$A_{1,CP}^{FB}$ can be much larger than $C\!P$As corresponding to the two-body decays embedded in the first two terms in Eq. (\ref{eq:Msqr}), provided that $\frac{a_{N_1}^{\text{tree}}}{a_{N_2}^{\text{tree}}}\neq \frac{a_{N_1}^{\text{penguin}}}{a_{N_2}^{\text{penguin}}}$.\footnote{If $\frac{a_{N_1}^{\text{tree}}}{a_{N_2}^{\text{tree}}}= \frac{a_{N_1}^{\text{penguin}}}{a_{N_2}^{\text{penguin}}}$, the last term of Eq. (\ref{eq:Msqr}) can be expressed as $|\alpha|^2\Re\left(\frac{c e^{i\delta}}{s_{N_1}s_{N_2}^\ast}\right)\cos\theta$, where $\alpha$ can be either $\alpha_{N_1}$ or $\alpha_{N_2}$. As a result, the strong phase $\delta$ is isolated from the $C\!P$V-related factor $|\alpha|^2$ and does not affect the $C\!P$V at all. It should be pointed out that this problem arises if one adopts the naive factorization approach for the weak decay processes of the current decay mode. This does not mean that the FB-$C\!P$A $A_{1,CP}^{FB}$ is small.
Rather, it indicates that one must go beyond the naive factorization approach for the study of the FB-$C\!P$A.}

\section{Summary}
%{\it Summary.---}
In summary, a set of angular distribution asymmetry observables, which are called the $j$-th forward-backward asymmetry for odd $j$, are introduced.
They can be used in searching for $C\!P$ violations in decay channels of bottom baryons.
Two typical situations for the application of the newly introduced observables were discussed.
The first is the two-body decay of $\Lambda_b^0\to \Lambda^0 M$, $\Lambda_b^0\to \Sigma^\pm M$, and $\Lambda_b^0\to \Lambda_c^+ M$, with $\Lambda^0$, $\Sigma^\pm$, or $\Lambda_c^+$ decaying weakly to two hadrons.
In this situation, the newly introduced observables are in fact equivalent to the decay asymmetry parameters.
The second situation corresponds to three-body decays of bottom baryons with the interference of intermediate resonances of similar masses and opposite parities.
A typical example for the second situation is the decay channel $\Xi_b^-\to p K^- K^-$, where possible interference between intermediate resonances $\Sigma(1775)$ and $\Sigma(1915)$ is present.
We suggest measuring $A^{FB}_1$, $A^{FB}_3$, and $A^{FB}_5$, and the corresponding $C\!P$ asymmetry observables $A^{FB}_{1,C\!P}$, $A^{FB}_{3,C\!P}$, and $A^{FB}_{5,C\!P}$.
Other examples include $\Lambda_b^0\to p K_S\pi^-$, $\Lambda_b^0\to \Lambda^0\pi^+\pi^-$, and so on.
We also use the decay $\Lambda_b^0\to p \pi^0 \pi^-$ to illustrate the basic idea.
Last but not least, the measurements of the angular distribution asymmetry observables and their corresponding $C\!P$ violation observables can also be performed in other decay channels of bottom or charmed hadrons.

\begin{acknowledgments}
One of the authors (Z.H.Z.) would like to thank Chia-Wei Liu and Chao-Qiang Geng for helpful discussions.
This work was supported by the National Natural Science Foundation of China (No. 12192261), Natural Science Foundation of
Hunan Province (No. 2022JJ30483), Natural Science Foundation of Zhejiang Province (No. LQ21A050005), and Ningbo Natural Science Foundation (No. 2021J180).
\end{acknowledgments}

\appendix
\section{Decay amplitude for $H\to h_1h_2h_3$}
%{\it Decay amplitude for $H\to h_1h_2h_3$.---}
To obtain the expression of $w^{(j)}$ in Eq. (\ref{wj}), one needs to write down the decay amplitude for the cascade decay $H\to R_i h_3$, $R_i\to h_1h_2$.
Two reference frames, the rest frame of $H$ (RF$H$) and that of $R_i$ (RF$R_i$), are needed for the helicity forms of the decay amplitudes.
In RF$H$, the $z$-axis is chosen along the quantization direction of the spin of $H$, and the momenta (helicities) of $R_i$ and $h_k$ ($k=1,2,3$), are denoted as $p$($\sigma$) and $q_k$($\lambda_k$), respectively.
The decay amplitude of $H\to R_i h_3$ can be expressed in the helicity form as
\begin{equation}
\mathcal{M}_{\sigma \lambda_3}^{H,Jm_z}=\mathcal{F}^{J}_{R_i,\sigma\lambda_3} D^{J*}_{m_z,\sigma-\lambda_3}(\phi_{B},\theta_{B},0),
\end{equation}
where $D$ is the Wigner-$D$ matrix, $(\phi_{B},\theta_{B})$ are the polar and azimuthal angles of $\vec{p}$ in RF$H$, $\mathcal{F}$ is the helicity decay amplitude.
In RF$R_i$, the $z'$-axis is chosen along the direction of the three-momentum of $H$ \cite{Gottfried:1964nx}.
The reason for choosing the $z'$-axis this way is that the helicity of $R_i$ in the RF$H$ is just the $z'$-component of the spin of $R_i$ in RF$R_i$.
In RF$R_i$, the momenta (helicities) of $h_k$ ($k=1,2,3$), will be denoted as  $q'_k$($\lambda'_k$).
The decay amplitude for $R_i\to h_1h_2$ can be expressed as
\begin{equation}
\mathcal{M}_{\lambda_1'\lambda_2'}^{R_i,s_{R_i}\sigma}=\mathcal{F}^{R_i,s_{R_i}}_{\lambda_1'\lambda_2'} D^{s*}_{\sigma,\lambda_1'-\lambda_2'}(\phi_1',\theta_1',0),
\end{equation}
where $(\phi_1',\theta_1')$ is the polar and azimuthal angles of $\vec{q}_{1} ^{~\prime}$ in RF$R_i$.
The decay amplitude for $H\to h_1h_2h_3$ (with $h_1h_2$ decaying from $R_i$'s) can then be expressed as
\begin{equation}
\mathcal{M}_{\lambda_1\lambda_2\lambda_3}^{Jm_z}=\sum_i \frac{\sum_{\sigma}\tilde{\mathcal{M}}_{\lambda_1\lambda_2}^{R_i,s_{R_i}\sigma}\mathcal{M}_{\sigma\lambda_3}^{H,Jm_z}}{\mathcal{I}_{R_i}},
\end{equation}
where $\mathcal{I}_{R_i}={s_{12}-m_{R_i}^2+i m_{R_i}\Gamma_{R_i}}$, $s_{jk}=(q_j+q_k)^2$,  ($j,k=1,2,3$) is the invariant mass squared of $h_j$ and $h_k$, $\Gamma_{R_i}$ is the decay width of $R_i$, and $\tilde{\mathcal{M}}_{\lambda_1\lambda_2}^{R_i,s_{R_i}\sigma}$ is the decay amplitude of $R_i\to h_1h_2$ in RF$H$, which can be obtained by a Lorentz transform of the amplitude $\mathcal{M}_{\lambda_1'\lambda_2'}^{R_i,s_{R_i}\sigma}$ according to
\begin{equation}
\tilde{\mathcal{M}}_{\lambda_1\lambda_2}^{R_i,s_{R_i}\sigma}=\sum_{\lambda_1'\lambda_2'}
\mathcal{M}_{\lambda_1'\lambda_2'}^{R_i,s_{R_i}\sigma}D^{j_1*}_{\lambda_1'\lambda_1}(\phi_{W1}, \theta_{W1},0)D^{j_2*}_{\lambda_2'\lambda_2}(\phi_{W2}, \theta_{W2},0),
\end{equation}
with $\Omega_{Wk}=(\phi_{W_k}, \theta_{W_k})$, ($k=1,2$) being the polar and azimuthal angles of the Wigner rotation, and $W_k$ being a pure Lorentz boost that transforms $q_k$ into $q_k'$.

\bibliography{zzhbib}

\end{document}